\documentclass[runningheads]{llncs}
\usepackage[T1]{fontenc}
\usepackage{graphicx}
\usepackage{amssymb}
\usepackage{lipsum}

\usepackage{adjustbox}
\usepackage{multirow}
\usepackage{wrapfig}
\usepackage{subfig}
\usepackage{tikz}
\usepackage{comment}
\usepackage{amsmath,amssymb}
\usepackage{hyperref}
\usepackage{algorithm}
\usepackage{xcolor}
\usepackage{listings}
\usepackage{stmaryrd}
\usepackage{colortbl}
\usepackage{booktabs} 
\usepackage{nicefrac}
\definecolor{mygray}{gray}{.9}
\usepackage{color}
\usepackage{bm}
\usepackage{wasysym}

\begin{document}

\title{Cross-modulated Few-shot Image Generation for Colorectal Tissue Classification}
\titlerunning{Few-shot Image Generation for Colorectal Tissue Classification}

\author{Amandeep Kumar \inst{1} \and
Ankan kumar Bhunia \inst{1} \and
Sanath Narayan\inst{3} \and
Hisham Cholakkal\inst{1} \and
Rao Muhammad Anwer \inst{1,2} \and
Jorma Laaksonen \inst{2} \and \\
Fahad Shahbaz Khan \inst{1}
}
\authorrunning{Kumar et al.}

\institute{$^1$MBZUAI, UAE, $^2$Aalto University, Finland \\ 
$^3$ Technology Innovation Institute, UAE \\
\email{amandeep.kumar@mbzuai.ac.ae}}

%


\maketitle              
\begin{abstract}
{In this work, we propose a few-shot colorectal tissue image generation method for addressing the scarcity of histopathological training data for rare cancer tissues.  Our  few-shot generation method, named XM-GAN, takes one base and a pair of reference tissue images as input and  generates high-quality yet diverse images. Within our XM-GAN, a novel controllable fusion block densely aggregates local regions of reference images based on their similarity to those in the base image, resulting in locally consistent features. To the best of our knowledge, we are the first to investigate few-shot generation in colorectal tissue images. We evaluate our few-shot colorectral tissue image generation by performing extensive qualitative, quantitative and subject specialist (pathologist) based evaluations. Specifically, in specialist-based evaluation, pathologists could differentiate between our XM-GAN generated tissue images 
 and real images only  $55\%$ time.  
Moreover, we utilize these generated images as data augmentation to address the few-shot tissue image classification task, achieving a gain of 4.4\% in terms of mean accuracy over the vanilla few-shot classifier.
Code: \url{https://github.com/VIROBO-15/XM-GAN} .}


\end{abstract}

\section{Introduction}
Histopathological image analysis is an important  step towards  cancer diagnosis.  However, shortage of pathologists worldwide along with the complexity of histopathological data make this task time consuming and challenging. Therefore, developing automatic and accurate histopathological image analysis methods that leverage recent progress in deep learning has received significant attention in recent years.  In this work, we investigate the problem of diagnosing colorectal cancer, which is one of the most common reason for cancer deaths around the world and particularly in Europe and America \cite{yu2021accurate}.

Existing deep learning-based colorectal tissue classification methods~\cite{wang2017histopathological,ohata2021novel,wang2021accurate} typically require large amounts of annotated histopathological training data for all tissue types to be categorized. However, obtaining large amount of training data is challenging, especially for rare cancer tissues.  To this end, it is desirable to develop a few-shot colorectal tissue classification method, which can learn from seen tissue classes having sufficient training data, and be able to transfer this knowledge to \textit{unseen} (novel) tissue classes having only  a \textit{few} exemplar training images. 

While generative adversarial networks (GANs) \cite{goodfellow2014generative} have been utilized to synthesize  images, they typically need to be trained using large amount of real  images of the respective classes, which is not feasible in aforementioned few-shot setting.  Therefore, we propose a few-shot (FS) image generation approach for generating high-quality and diverse colorectal tissue images of novel classes using limited exemplars. Moreover, we demonstrate the applicability of these generated images for the challenging problem of FS colorectal tissue classification.  

\noindent\textbf{Contributions:} 
We propose a few-shot colorectal tissue image generation framework, named XM-GAN, which simultaneously focuses on generating high-quality yet diverse images. Within our tissue image generation framework, we introduce a novel controllable fusion block (CFB) that enables a dense aggregation of local regions of the reference tissue images based on their congruence to those in the base tissue image. Our CFB employs a cross-attention based feature aggregation between the base (\textit{query}) and reference (\textit{keys}, \textit{values}) tissue image features. Such a cross-attention mechanism enables the aggregation of reference features from a global receptive field, resulting in \textit{locally} consistent features. Consequently, colorectal tissue images are generated with reduced artifacts.

\begin{figure}[t!]
\centering
\includegraphics[width=0.90\textwidth]{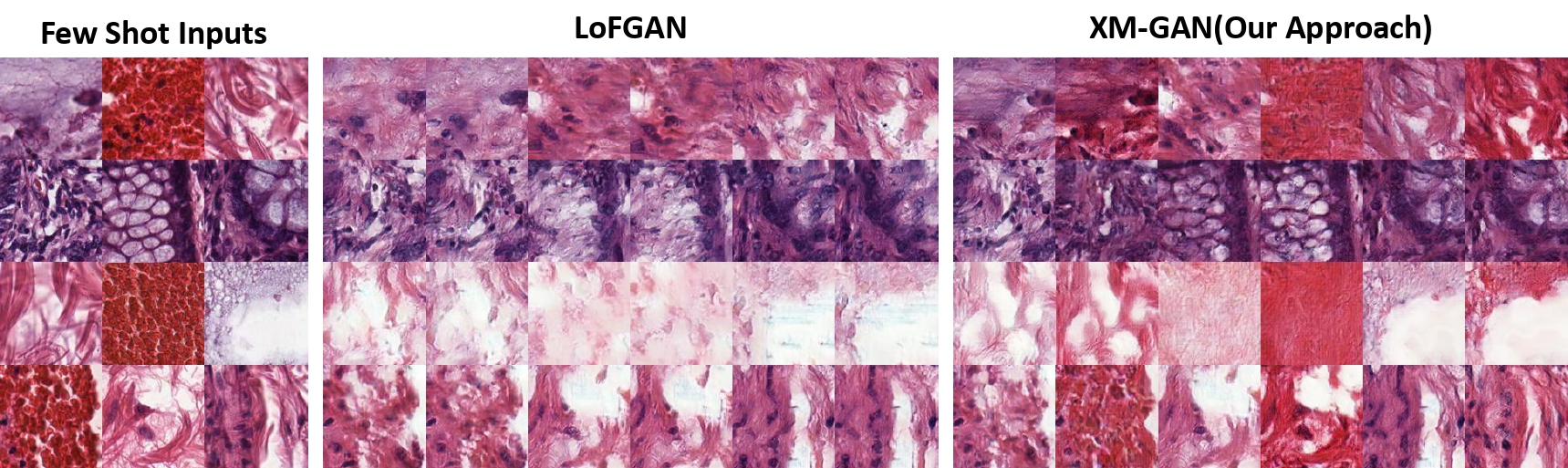} \vspace{-0.3cm}
\caption{On the left: few-shot input images of colorectal tissues. In the middle: images generated by LoFGAN. On the right: images generated by our XM-GAN. Compared to LoFGAN, our XM-GAN generates images that are high-quality yet diverse. Best viewed zoomed in. Additional results are provided in the supplementary material.}
\label{fig:qual}
\vspace{-0.4cm}
\end{figure}

To further enhance the diversity and quality of the generated tissue images, we introduce a mapping network along with a controllable cross-modulated layer normalization (cLN) within our CFB. Our mapping network generates `meta-weights' that are a function of the global-level features of the reference tissue image and the control parameters. These meta-weights are then used to compute the modulation weights for feature re-weighting in our cLN. This enables the cross-attended tissue image features to be re-weighted and enriched in a controllable manner, based on the reference tissue image features and associated control parameters. Consequently, it results in improved diversity of the tissue images generated by our transformer-based framework (see Fig.~\ref{fig:qual}).

We  validate our XM-GAN on the FS colorectral tissue image generation task by performing extensive qualitative, quantitative and subject specialist (pathologist) based evaluations. Our XM-GAN generates realistic \textit{and} diverse colorectal tissue images (see Fig.~\ref{fig:qual}). In our subject specialist (pathologist) based evaluation, pathologists could differentiate between our XM-GAN generated colorectral tissue images 
 and  real images \textit{only}  $55\%$ time.  
Furthermore, we evaluate the effectiveness of our generated tissue images by using them as data augmentation during  training of FS colorectal tissue image classifier, leading to an absolute gain of $4.4\%$ in terms of mean classification accuracy over the vanilla FS classifier.

\section{Related Work}

The ability of generative models~\cite{goodfellow2014generative,kingma2013auto} and transformer~\cite{thawakar20233d,shamshad2022transformers}  to fit to a variety of data distributions has enabled great strides of advancement in tasks, such as image generation~\cite{karras2017progressive,karras2019style,brock2018large,vahdat2020nvae,kumar2021udbnet}, and so on. Despite their success, these generative models typically require large amount of data to train and avoid overfitting. In contrast, few-shot (FS) image generation approaches~\cite{clouatre2019figr,liang2020dawson,bartunov2018few,hong2020matchinggan,gu2021lofgan} strive to generate natural images from disjoint novel categories from the same domain as in the training. Existing FS natural image generation approaches can be broadly divided into three categories based on transformation \cite{antoniou2017data}, optimization \cite{clouatre2019figr,liang2020dawson}  and fusion \cite{hong2020matchinggan,hong2020f2gan,gu2021lofgan}. The transformation-based approach learns to perform generalized data augmentations to generate intra-class images  from a single conditional image. On the other hand, optimization-based approaches typically utilize meta-learning techniques to adapt to a different image generation task by optimizing on a few reference images from the novel domain. Different from these two paradigms that are better suited for simple image generation task, fusion-based approaches first aggregate latent features of reference images and then employ a decoder to generate same class images from these aggregated features.

\noindent\textbf{Our Approach:}
While the aforementioned works explore FS generation in \textit{natural} images, to the best of our knowledge, we are the first to investigate \textit{FS generation in colorectal tissue} images. 
In this work, we look into multi-class colorectal tissue analysis problem, with low and high-grade tumors included in the set. The corresponding dataset \cite{kather_2016_53169} used in this study is widely employed for multi-class texture classification in colorectal cancer histology and comprises eight types of tissue: tumor epithelium, simple stroma, complex stroma, immune cells, debris, normal mucosal glands, adipose tissue and background (no tissue). Generating colorectal tissue images of these diverse categories is a challenging task, especially in the FS setting.  Generating realistic and diverse tissue images require ensuring both global and local texture consistency (patterns). Our XM-GAN densely aggregates features~\cite{vaswani2017attention,dosovitskiy2020image} from all relevant local regions of the reference tissue images at a global-receptive field along with a controllable mechanism for modulating the tissue image features by utilizing meta-weights computed from the input reference tissue image features. As a result, this leads to high-quality yet diverse colorectal tissue image generation in FS setting.

\section{Method}

\noindent\textbf{Problem Formulation:} In our few-shot colorectal tissue image generation framework, the goal is to generate diverse set of images from $K$ input examples $X$ of a \textit{unseen} (novel) tissue classes. Let $\mathcal{D}^{s}$ and $\mathcal{D}^{u}$ be the set of seen and unseen classes, respectively, where $\mathcal{D}^{s} \cap \mathcal{D}^{u} = \emptyset$. In the training stage, we sample images from $\mathcal{D}^{s}$ and train the model to learn transferable generation ability to produce new tissue images for unseen classes. During inference, given $K$ images from an unseen class in $\mathcal{D}^{u}$, the trained model strives to produce diverse yet plausible images for this unseen class without any further fine-tuning.

\begin{figure}[t!]
\centering
\includegraphics[width=\textwidth]{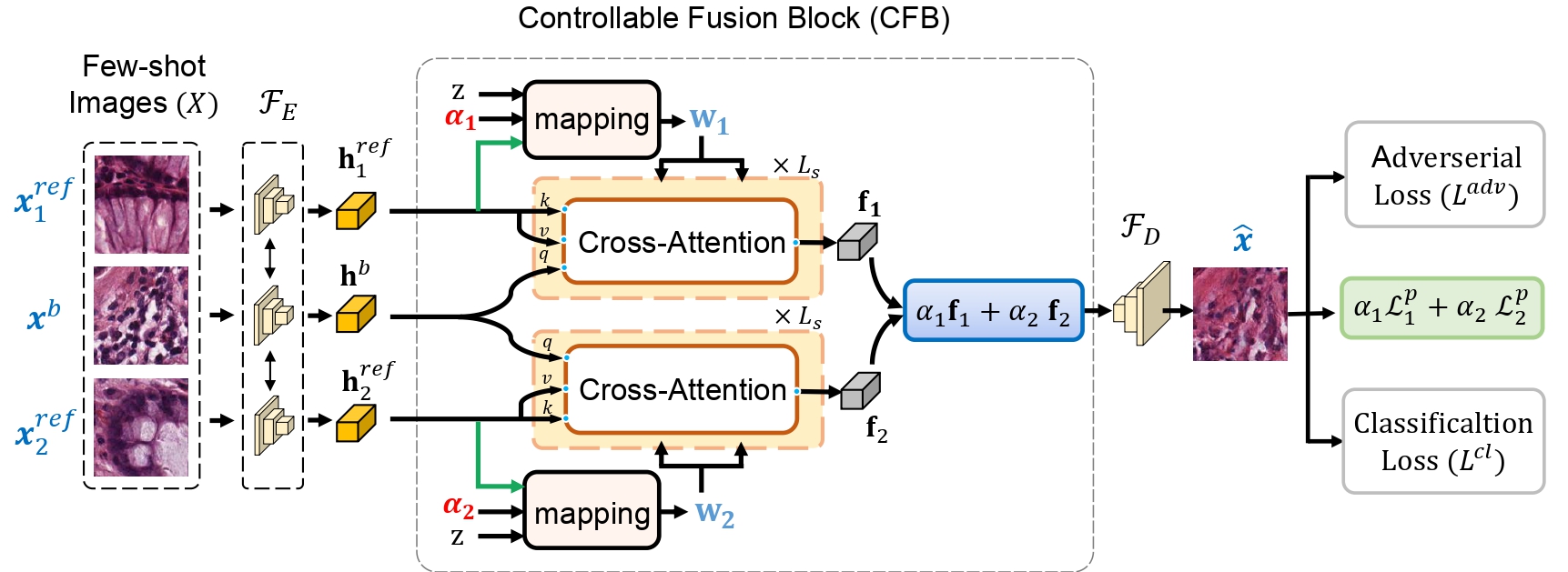} \vspace{-0.4cm}
\caption{Our XM-GAN comprises a CNN encoder, a transformer-based controllable fusion block (CFB), and a CNN decoder for tissue image generation. For $K$-shot setting, a shared encoder $\mathcal{F}_E$ takes a base tissue image $x^b$ along with $K{-}1$ reference tissue images $\{x_i^{ref}\}_{i=1}^{K-1}$ and outputs visual features $\mathbf{h}^b$ and $\{\mathbf{h}_i^{ref}\}_{i=1}^{K-1}$, respectively. Within our CFB, a mapping network computes meta-weights $\mathbf{w}_i$ which are utilized to generate the modulation weights for feature re-weighting during cross-attention. The cross-attended features $\mathbf{f}_i$ are fused and input to a decoder $\mathcal{F}_D$ that generates an image $\hat{x}$. 
} \vspace{-0.3cm}
\label{fig:main}
\end{figure}

\noindent\textbf{Overall Architecture:} Fig.~\ref{fig:main} shows the overall architecture of our proposed framework, XM-GAN. Here, we randomly assign a tissue image from $X$ as a base image $x^{b}$, and denote the remaining $K{-}1$ tissue images
as reference $\{x_i^{ref}\}_{i=1}^{K-1}$. Given the input images $X$, we obtain feature representation of the base tissue image and each reference tissue image by passing them through the shared encoder $\mathcal{F}_E$. Next, the encoded feature representations $\mathbf{h}$ are input to a controllable fusion block (CFB), where cross-attention~\cite{vaswani2017attention} is performed between the base and reference features, $\mathbf{h}^b$ and $\mathbf{h}^{ref}_i$, respectively. Within our CFB, we introduce a mapping network along with a controllable cross-modulated layer normalization (cLN) to compute meta-weights $\mathbf{w}_i$, which are then used to generate the modulation weights used for re-weighting in our cLN. The resulting fused representation $\mathbf{f}$ is input to a decoder $\mathcal{F}_D$ to generate tissue image $\hat{x}$. The whole framework is trained following the GAN~\cite{lim2017geometric} paradigm. In addition to $\mathcal{L}^{adv}$ and $\mathcal{L}^{cl}$, we propose to use a guided perceptual loss term $\mathcal{L}^p$, utilizing the control parameters $\alpha_i$. Next, we describe our CFB in detail.

\subsection{Controllable Fusion Block\label{sec:CFB}}
\begin{wrapfigure}{r}{0.4\textwidth} 
  \begin{center}
  \vspace{-0.9cm}
 \includegraphics[width=0.35\textwidth]{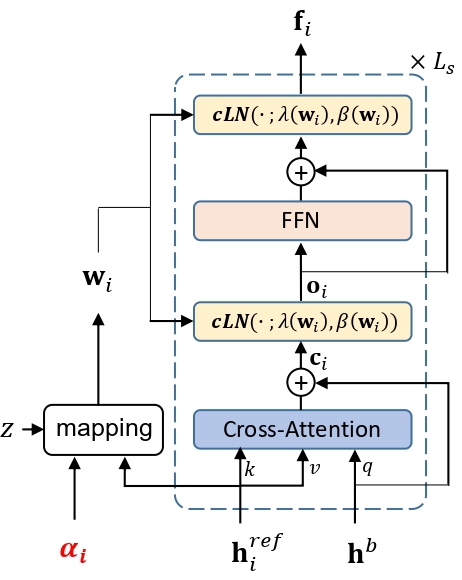}\vspace{-0.3cm}
    \caption{Cross-attending the base and reference tissue image features using controllable cross-modulated layer norm (cLN) in our CFB. Here, a reference feature $\mathbf{h}_i^{ref}$, noise $\bm{z}$ and control parameter $\alpha_i$ are input to a mapping network for generating meta-weights $\mathbf{w}_i$. The resulting $\mathbf{w}_i$ modulates the features via $\lambda({\mathbf{w}_i})$ and $\beta({\mathbf{w}_i})$ in our cLN. As a result of this controllable feature modulation, the output features $\mathbf{f}_i$ enable the generation of tissue images that are diverse yet aligned with the semantics of the input tissue images.} \vspace{-0.3cm}
    \label{fig:sln}
  \end{center}
\end{wrapfigure}

Fig.~\ref{fig:sln} shows the architecture of our proposed CFB, comprises of a shared cross-transformer followed by a feature fusion mechanism. Here, the cross-transformer is based on multi-headed cross-attention mechanism that densely aggregates relevant input image features, based on pairwise attention scores between each position in the base tissue image with every region of the reference tissue image. The \textit{query} embeddings $\bm{q}^m \in \mathbb{R}^{n \times d }$ are computed from the base features $\mathbf{h}^b \in \mathbb{R}^{n\times D}$, while \textit{keys} $\bm{k}_i^m \in \mathbb{R}^{n \times d}$ and \textit{values} $\bm{v}_i^m \in \mathbb{R}^{n \times d}$ are obtained from the reference features $\mathbf{h}^{ref}_i \in \mathbb{R}^{n\times D}$, where $d=\nicefrac{D}{M}$ with $M$ as the number of attention heads. Next, a cross-attention function maps the queries to outputs $\bm{r}_i^m$ using the key-value pairs. Finally, the outputs $\bm{r}_i^m$ from all $M$ heads are concatenated and processed by a learnable weight matrix $\mathbf{W}\in \mathbb{R}^{D\times D}$ to generate cross-attended features $\bm{c}_i  \in \mathbb{R}^{n\times D}$ given by
\begin{equation}
\bm{c}_i = [\bm{r}_i^1; \cdots; \bm{r}_i^M]\mathbf{W} + \mathbf{h}^b, \qquad \text{where} \quad \bm{r}_i^m = \text{softmax}(\frac{\bm{q}^m\bm{k}_i^{m\top}}{\sqrt{d}})\bm{v}_i^m.
\end{equation}

Next, we introduce a controllable feature modulation mechanism in our cross-transformer to further enhance the diversity and quality of generated images.

\noindent\textbf{Controllable Feature Modulation:} The standard cross-attention mechanism, described above, computes locally consistent features that generate images with reduced artifacts. However, given the deterministic nature of the cross-attention and the limited set of reference images, simultaneously generating diverse and high-quality images in the few-shot setting is still a challenge. 
To this end, we introduce a controllable feature modulation mechanism within our CFB that aims at improving the  diversity and quality of generated images. The proposed modulation incorporates stochasticity as well as enhanced control in the feature aggregation and refinement steps. This is achieved by utilizing the output of a mapping network for modulating the visual features in the layer normalization modules in our cross-transformer.

\noindent\textit{Mapping network:} The meta-weights $\mathbf{w}_i \in \mathbb{R}^D$ are obtained by the mapping network as, 
\begin{equation}
    \label{eqn_map}
    \mathbf{w}_i = \mathbf{g}^{ref}_i \odot \psi_\alpha(\alpha_i) + \psi_z(\bm{z}),
\end{equation}where $\psi_\alpha(\cdot)$ and $\psi_z(\cdot)$ are linear transformations, $\bm{z} \sim \mathcal{N}(0,1)$ is a Gaussian noise vector, and $\alpha_i$ is control parameter. $\mathbf{g}_i^{ref}$ is global-level feature computed from the reference features $\mathbf{h}_i^{ref}$ through a linear transformation and a global average pooling operation. The meta-weights $\mathbf{w}_i$ are then used for modulating the features in our cross-modulated layer normalization, as described below.

\noindent\textit{Controllable Cross-modulated Layer Normalization (cLN):} Our cLN learns \textit{sample-dependent} modulation weights for normalizing features since it is desired to generate images that are similar to the few-shot samples. Such a dynamic modulation of features enables our framework to generate images of high-quality and diversity. To this end, we utilize the meta-weights $\mathbf{w}_i$ for computing the modulation parameters $\lambda$ and $\beta$ in our layer normalization modules. With the cross-attended feature $\mathbf{c}_i$ as input, our cLN modulates the input to produce an output feature $\mathbf{o}_i \in \mathbb{R}^{n\times D}$, given by
\begin{equation}
    \mathbf{o}_i = \text{cLN}(\mathbf{c}_i, \mathbf{w}_i) = \lambda(\mathbf{w}_i) \odot \frac{\mathbf{c}_i-\mu}{\sigma} + \beta(\mathbf{w}_i),
\end{equation}
where $\mu$ and $\sigma^2$ are the estimated mean and variance of the input $\mathbf{c}_i$. Here, $\lambda(\mathbf{w}_i)$ is computed as the element-wise multiplication between meta-weights $\mathbf{w}_i$ and sample-independent learnable weights $\bm{\lambda} \in\mathbb{R}^D$, as $\bm{\lambda} \odot \mathbf{w}_i$. A similar computation is performed for $\beta(\mathbf{w}_i)$. Consequently, our proposed normalization mechanism achieves a controllable modulation of the input features based on the reference image inputs and enables enhanced diversity and quality in the generated images. The resulting features $\mathbf{o}_i$ are then passed through a feed-forward network (FFN) followed by another cLN for preforming point-wise feature refinement, as shown in Fig.~\ref{fig:sln}. Afterwards, the cross-attented features $\mathbf{f}_i$ are aggregated using control parameters $\alpha_i$ to obtain the fused feature representation $\mathbf{f} = \sum_i \alpha_i \mathbf{f}_i$, where $i\in[1,\cdots,K-1]$. Finally, the decoder $\mathcal{F}_D$ generates the final image $\hat{x}$.

\subsection{Training and Inference}
\textbf{Training:} The whole framework is trained end-to-end following the hinge version GAN~\cite{lim2017geometric} formulation. With generator $\mathcal{F}_G$ denoting our encoder, CFB and decoder together, and discriminator $\mathcal{F}_{Dis}$, the adversarial loss $\mathcal{L}^{adv}$ is given by
\begin{equation}
\begin{aligned}
\begin{gathered}
     \mathcal{L}^{adv}_{\mathcal{F}_{Dis}}= \underset{x \sim real}{\mathbb{E}}[\text{max}(0,1-\mathcal{F}_{Dis}(x))] + \underset{\hat{x} \sim fake}{\mathbb{E}}[\text{max}(0,1+\mathcal{F}_{Dis}(\hat{x}))] \\
     \text{and} \quad  \mathcal{L}^{adv}_{\mathcal{F}_G} = -\underset{\hat{x} \sim fake}{\mathbb{E}}[\mathcal{F}_{Dis}(\hat{x})]. 
     \end{gathered}
\end{aligned}
\end{equation}
Additionally, to encourage the generated image $\hat{x}$ to be perceptually similar to the reference images based on the specified control parameters $\bm{\alpha}$, we use a parameterized formulation of the standard perceptual loss \cite{johnson2016perceptual}, given by 
\begin{equation}
    \label{loss:perceptual}
    \mathcal{L}^p = \sum_i\alpha_i \mathcal{L}_i^p, \qquad \text{where} \quad \mathcal{L}^p_i = \mathbb{E}[\| \phi(\hat{x})-\phi(x_i^{ref}) \|_2].
\end{equation}
Moreover, a classification loss $\mathcal{L}^{cl}$ enforces that the images generated by the decoder are classified into the corresponding class of the input few-shot samples. Our  XM-GAN is then trained using the formulation:
%
%
    $\mathcal{L}   = \mathcal{L}^{adv} + \eta_{p}\mathcal{L}^p + \eta_{cl}\mathcal{L}^{cl}$,
where $\eta_p$ and $\eta_{cl}$ are hyperparameters for weighting the loss terms.

\noindent \textbf{Inference:} During inference, multiple high-quality and diverse images $\hat{x}$ are generated by varying the control parameter $\alpha_i$ for a set of fixed $K$-shot samples. 
While a base image $x^b$ and $\alpha_i$ can be randomly selected, our framework enables a user to have control over the generation based on the choice of $\alpha_i$ values.

\section{Experiments}

We conduct experiments on human colorectal cancer dataset \cite{kather_2016_53169}. The dataset consist of $8$ categories of colorectal tissues, Tumor, Stroma, Lymph, Complex, Debris, Mucosa, Adipose, and Empty with 625 per categories. To enable few-shot setting, we split the 8 categories into 5 seen (for training) and 3 unseen categories (for evaluation) with 40 images per category.  We evaluate our approach using two metrics: Fr\`echet Inception Distance (FID)~\cite{heusel2017gans} and Learned Perceptual Image Patch Similarity (LPIPS)~\cite{zhang2018unreasonable}. Our encoder $\mathcal{F}_E$ and decoder $\mathcal{F}_D$ both have five convolutional blocks with batch normalization and Leaky-ReLU activation, as in~\cite{gu2021lofgan}. The input and generated image image size is 128$\times$128. The linear transformation $\psi(\cdot)$ is implemented as a 1$\times$1 convolution with input and output channels set to $D$. The weights $\eta_{p}$  and $\eta_{cl}$ are set to 50 and 1. We set $K=3$ in all the experiments, unless specified otherwise. Our XM-GAN is trained with a batch-size of 8 using the Adam optimizer and a fixed learning rate of $10^{-4}$.

\subsection{State-of-the-art Comparison}

\noindent \textbf{FS Tissue Image Generation:} In Tab.~\ref{tab:main}, we compare our XM-GAN approach for FS tissue image generation with state-of-the-art LoFGAN~\cite{gu2021lofgan} on \cite{kather_2016_53169} dataset. Our proposed XM-GAN that utilizes dense aggregation of relevant local information at a global receptive field along with controllable feature modulation outperforms LoFGAN with a significant margin of $30.1$, achieving FID score of $55.8$. Furthermore, our XM-GAN achieves a better LPIPS score. In Fig.~\ref{fig:qual}, we present a qualitative comparison of our XM-GAN with  LoFGAN~\cite{gu2021lofgan}.

\begin{wraptable}{r}{5cm}
\vspace{-0.5cm}
\caption{Our XM-GAN achieves consistent gains in performance on both FID and LPIPS scores, outperforming LoFGAN on \cite{kather_2016_53169} dataset.}\vspace{-0.3cm}
\label{tab:main}
\scalebox{0.85}{
\begin{tabular}{l|cc}
\toprule[0.2mm]
\rowcolor{mygray} 
{\cellcolor{mygray}\textbf{Method}} &
  FID($\downarrow$) & LPIPS($\uparrow$) \\ \midrule
  
LoFGAN~\cite{gu2021lofgan}     &85.9 &0.44 \\
\midrule
\textbf{Ours: XM-GAN}          &\textbf{55.8} &\textbf{0.48}  \\  \bottomrule[0.2mm]
\end{tabular}\vspace{-0.4cm}
}
\end{wraptable} 

\noindent \textbf{Low-data Classification:}
Here, we evaluate the applicability of the tissue images generated by our XM-GAN as a source of data augmentation for the downstream task of low-data colorectal tissue classification for unseen categories. The unseen dataset is split into $D_{tr}$, $D_{val}$, $D_{test}$. Images of an unseen class are split into 10:15:15.
Following~\cite{gu2021lofgan}, seen categories are used for initializing the ResNet18 backbone and a new classifier is trained using $D_{tr}$. We refer to this as Standard. Then, we augment $D_{tr}$ with $30$ tissue images generated by our XM-GAN using the same $D_{tr}$ as few-shot samples for each unseen class. Tab.~\ref{tab:cls} shows the classification performance comparison. 
Compared to the LoFGAN~\cite{gu2021lofgan}, our XM-GAN achieves absolute gains of 2.8\%. 

\subsection{Ablation Study}
\label{sec:abl}
Here, we present our ablation study to validate the merits of the proposed contributions. Tab.~\ref{tab:abl} shows the baseline comparison on the \cite{kather_2016_53169} dataset. Our \texttt{Baseline} comprises an encoder, a standard cross-transformer with standard Layer normalization (LN) layers and a decoder. This is denoted as \texttt{Baseline}. \texttt{Baseline+PL} refers to extending
the \texttt{Baseline} by also integrating the standard perceptual loss.
We conduct an additional experiment using random values of $\alpha_i$ \textit{s.t.} $\sum_i \alpha_i {=}1$ for computing the fused feature $\mathbf{f}$ and parameterized perceptual loss (Eq.~\ref{loss:perceptual}). We refer to this as \texttt{Baseline+PPL}. Our final XM-GAN referred here as \texttt{Baseline+PPL+cLN} contains the novel CFB. Within our CFB, we also validate the impact of the reference features for feature modulation by computing the meta-weights $\mathbf{w}_i$ using \textit{only} the Gaussian noise $\bm{z}$ in Eq.~\ref{eqn_map}. This is denoted here as \texttt{Baseline+PPL+cLN}${}^\dagger$.  Our approach based on the novel CFB achieves the best performance amongst all baselines.


\begin{table}[t]
\parbox{.38\linewidth}{
\centering
\caption{Low-data image classification. The proposed XM-GAN achieves superior classification performance compared to recently introduced LoFGAN.} 
\scalebox{0.88}{
\label{tab:cls}
        \begin{tabular}{l|c}
        
        \toprule[0.4mm]
        
        \rowcolor{mygray} 
        {\cellcolor{mygray}\textbf{Method}} &
          Accuracy (\%)  \\ \midrule
          
        Standard      &68.1    \\
        LoFGAN~\cite{gu2021lofgan}      &69.7   \\
        \midrule
        \textbf{Ours: XM-GAN}          &\textbf{72.5}   \\  \bottomrule[0.4mm]
        \end{tabular}\vspace{-0.3cm}
}
}
\hfill
\parbox{.58\linewidth}{
\centering
\caption{Impact of integrating parameterized perceptual loss (PPL) and cLN to the baseline. Please refer to Sec.~\ref{sec:abl} for more details.} 
\scalebox{0.84}{
\label{tab:abl}
            \begin{tabular}{l|cc}
            
            \toprule[0.4mm]
            
            \rowcolor{mygray} 
            \rowcolor{mygray} 
            {\cellcolor{mygray}\textbf{Method}} &
              \textbf{FID}($\downarrow$) & \textbf{LPIPS}($\uparrow$) \\ \midrule
              
            \texttt{Baseline}      & 73.6 & 0.451  \\
            \texttt{Baseline + PL}    &69.2 & 0.467  \\
            
            \texttt{Baseline + PPL}       & 66.5  & 0.471 \\
            \midrule
            \texttt{Baseline + PPL + cLN}${}^{\dagger}$       & 62.1 & 0.475   \\
            \textbf{Ours: Baseline + PPL + cLN}   & \textbf{55.8} & \textbf{0.482} 
              \\  \bottomrule[0.4mm]
            \end{tabular} \vspace{-0.3cm}
}
}
\end{table}

\subsection{Human Evaluation Study}
We conducted a study with a group of ten pathologists having an average subject experience of 8.5 years. Each pathologist is shown a random set of 20 images (10 real and 10 XM-GAN generated) and asked to identify whether they are real or generated. The study shows that pathologists could differentiate between the AI-generated and  real images \textit{only}  $55\%$ time, which is comparable with a random prediction in a binary classification problem, indicating the ability of our proposed generative framework to generate realistic colorectal images.

\section{Conclusions}

We proposed a few-shot colorectal tissue image generation approach that comprises a controllable fusion block (CFB) which generates locally consistent features by performing a dense aggregation of local regions from reference tissue images based on their similarity to those in the base tissue image. We introduced a mapping network together with a cross-modulated layer normalization, within our CFB, to enhance the quality and diversity of generated images. We extensively validated our XM-GAN by performing quantitative, qualitative and human-based evaluations, achieving state-of-the-art results.\\

\textbf{Acknowledgement}: We extend our heartfelt appreciation to the pathologists who made significant contributions to our project. We are immensely grateful to Dr. Hima Abdurahiman from Government Medical College-Kozhikode, India; Dr. Sajna PV from MVR Cancer Center and Research Institute, Kozhikode, India; Dr. Binit Kumar Khandelia from North Devon District Hospital ,UK ; Dr. Nishath PV from Aster Mother Hospital Kozhikode, India; Dr. Mithila Mohan from Dr. Girija's Diagnostic Laboratory and Scans, Trivandrum, India; Dr. Kavitha from Aster MIMS, Kozhikode, India; and several other unnamed pathologists who provided their expert advice, valuable suggestions, and insightful feedback throughout various stages of our research work.

\bibliographystyle{splncs04}
\bibliography{egbib}
\end{document}